\pdfoutput=1
\documentclass[english,article,prl,reprint,microtype,longbibliography,textcomp]{revtex4-1}
\usepackage[T1]{fontenc}
\usepackage[latin9]{inputenc}
\usepackage{color}
\usepackage{babel}
\usepackage{amstext}
\usepackage{amssymb}
\usepackage{graphicx}
\usepackage{subscript}
\usepackage[unicode=true,
 bookmarks=true,bookmarksnumbered=false,bookmarksopen=false,
 breaklinks=false,pdfborder={0 0 0},backref=false,colorlinks=true]
 {hyperref}
\hypersetup{
 pdfauthor={Ambroise van Roekeghem}}

\makeatletter
\date{}

\makeatother

\begin{document}

\title{Anomalous thermal conductivity\\and suppression of negative thermal
expansion in ScF\textsubscript{3}}

\author{Ambroise van Roekeghem, Jes\'{u}s Carrete and Natalio Mingo}

\affiliation{CEA, LITEN, 17 Rue des Martyrs, 38054 Grenoble, France}

\email{ambroise.van-roekeghem@polytechnique.edu}

\date{\today}
\begin{abstract}
The empty perovskite ScF$_{3}$ exhibits negative thermal expansion
up to 1100\,K. We demonstrate that \textit{ab-initio} calculations
of temperature-dependent effective phonon spectra allow to quantitatively
describe the behavior of this compound and the suppression of negative
thermal expansion. Based on this result, we predict an anomalous temperature
dependence of the thermal conductivity and interpret it as a general
feature of the perovskite class. Finally, we comment on the fact that
the suppression of negative thermal expansion at such a high temperature
is beyond the reach of the quasi-harmonic approximation and we discuss
this suppression based on the temperature-dependence of the mode Gr\"{u}neisen
parameters.
\end{abstract}
\maketitle
Negative thermal expansion (NTE), caused by anharmonicity in solids,
is crucial for technological applications to obtain materials with
no volume variation over given temperature ranges \cite{Dove_NTE_review}.
Being able to find new materials exhibiting this behavior from \textit{ab-initio}
calculations is thus an important issue. Recently, attention has been
drawn to fluorides and to empty perovskites, following the discovery
of ScF$_{3}$, which exhibits NTE up to 1100\,K \cite{Greve_ScF3_2010}.
Few attempts to understand and reproduce this behavior from realistic
calculations have been performed so far. Within the quasi-harmonic
approximation, contradictory results have been obtained by different
groups: while Li and collaborators found a persistent NTE at high
temperature \cite{Li_ScF3_quartic}, Liu and collaborators found the
suppression of NTE around 400\,K \cite{Liu_ScF3_ReO3}. In both cases,
the calculated lattice parameter was different from the experimental
measurement by about 1\% \cite{Li_Thesis_2012}, whereas the maximum
experimental variation of the lattice parameter due to the thermal
expansion is about 0.5\% \cite{Greve_ScF3_2010}. Moreover, the value
of the coefficient of thermal expansion (CTE) at low temperature was
also too small by at least a factor of two. Li and collaborators pointed
out that the structural quarticity of the transverse vibrations of
the fluorine atoms improves the behavior of the calculated NTE coefficient
with respect to experiment, with the NTE being stronger at low temperature
and weaker at high temperature as compared to the quasi-harmonic approximation.
Still, they could not perform a quantitative calculation over the
whole Brillouin zone and their approximation to the effect of quartic
anharmonicity failed to show the suppression of NTE at high temperature
\cite{Li_ScF3_quartic}. In contrast, calculations using density-functional
molecular dynamics, which take into account anharmonicity, reproduce
the suppression of NTE \cite{Lazar_molecular_dynamics_ScF3}. Besides
these difficulties, other clues highlight the importance of anharmonicity
in this material, and in the class of perovskites in general. In particular,
several measurements of the thermal conductivity in various perovskites
describe an anomalous temperature dependence and point out the role
of the soft modes in those compounds \cite{Tachibana_perovskite_kappa,Suemune_KZnF3,Martin_KZnF3_RbCaF3_KMgF3_KMnF3,Mante_BaTiO3,Muta_SrTiO3_kappa,Popuri_SrTiO3_kappa,Maekawa_BaHfO3_SrHfO3,Yamanaka_BaZrO3_kappa}.
So far, there is no calculation or measurement of the thermal conductivity
of ScF$_{3}$ available in the literature.

In this Letter, we use \textit{ab-initio} calculations of temperature-dependent
effective phonon dispersions and scattering rates to investigate the
effects of anharmonicity in ScF$_{3}$. We compute the temperature-dependent
interatomic force constants using regression analysis of forces from
density functional theory coupled with a harmonic model of the quantum
canonical ensemble, to obtain the evolution of the lattice parameter
and of the phonon spectrum with quantitative agreement with experimental
data up to 1000\,K. We then calculate the thermal conductivity and
predict an anomalous temperature dependence similar to what has been
measured in other perovskites, which is caused by anharmonicity. Finally,
we show that the suppression of negative thermal expansion at high
temperature is beyond the reach of the quasi-harmonic approximation,
and we discuss this phenomenon in terms of the mode Grüneisen parameters.

Several methods have been recently developed to deal with anharmonic
effects at finite temperature in solids. The Self-Consistent \textit{Ab-Initio}
Lattice Dynamical method (SCAILD) \cite{Souvatzis_SCAILD} establishes
a self-consistency loop on the phonon frequencies in the high-temperature
limit by computing forces created by displacements with an amplitude
fixed by the classical mean-square displacement. The Temperature-Dependent
Effective Potential approach (TDEP) \cite{Hellman_TDEP_2011,Hellman_TDEP_2013}
is based on \textit{ab-initio} molecular dynamics (AIMD) and relies
on fitting the computed forces from the AIMD trajectory to a model
Hamiltonian, which can include 2nd or higher order terms. In the Stochastic
Self-Consistent Harmonic Approximation (SSCHA) \cite{Errea_SSCHA},
the free energy of an ensemble of harmonic oscillators in the real
potential is minimized with respect to the 2nd order force constants
and to the structure of the compound. Finally, an approach \cite{Tadano_SCPH}
based on Self-Consistent Phonon theory (SCPH) \cite{Werthamer_SCPH}
estimates the anharmonic self-energy using third and fourth order
force constants extracted by compressive sensing techniques on DFT
calculations \cite{Zhou_compressive_sensing}. In our case, we use
an approach inspired by the SCAILD method but using the full quantum
mean square thermal displacement matrix as in the SSCHA and allowing
the possibility to update the eigenvectors and to obtain higher order
force constants by fitting the forces in real space, as in TDEP. In
the harmonic approximation, the probability to find the system in
a configuration in which each ion $i$ is displaced in direction $\alpha$
by $u_{i\alpha}$ is $\rho_{h}\left(\left\{ u_{i\alpha}\right\} \right)\propto\exp\left(-\frac{1}{2}u^{T}\mathbf{\Sigma}^{-1}u\right)$,
with $\Sigma\left(i\alpha,j\beta\right)$ the quantum covariance for
atoms $i,j$ and directions $\alpha,\beta$: 
\[
\Sigma\left(i\alpha,j\beta\right)=\frac{\hbar}{2\sqrt{M_{i}M_{j}}}\sum_{m}\omega_{m}^{-1}\left(1+2n_{B}\left(\omega_{m}\right)\right)\epsilon_{mi\alpha}\epsilon_{mj\beta}^{*}
\]

where $M$ is the atomic mass, $\omega_{m}$ the phonon frequency
of mode $m$ (comprising both wavevector and branch degrees of freedom),
$\mathbf{\epsilon}_{m}$ the corresponding wavefunction and $n_{B}$
the Bose-Einstein distribution. From the phonon frequencies and eigenvectors
of the starting phonon spectrum, we compute the matrix $\mathbf{\Sigma}$
using the $\Gamma$ point of a 4x4x4 supercell and use it as the covariance
matrix of a multidimensional Gaussian distribution to generate $N$
random sets of atomic displacements $\left\{ u_{i\alpha}^{n}\right\} $,
with $i$ indexing each atom and direction in the supercell and $n$
a given random configuration. The forces $\mathbf{f}_{i}^{n}$ acting
on each atom of the supercell generated by this set of displacements
$\left\{ u_{i\alpha}^{n}\right\} $ are calculated by density functional
theory. Finally, we use those forces and displacements to fit the
second and third order forces constants \footnote{In this study, we use a cutoff of 5\,\AA{} for the third order force
constants.} of a model potential, using least-squares minimization. These force
constants allow us to calculate a new phonon spectrum. In practice,
we start from the spectrum calculated using small displacements and
we iterate the procedure described above until convergence. In addition,
for each cycle we calculate the external pressure $P=\left(\frac{\partial E_{p}}{\partial V}\right)_{S}+\left(\frac{\partial E_{k}}{\partial V}\right)_{S}$.
The derivative of the potential energy $E_{p}$ is computed using
the mean of the external pressure $P^{DFT}$ obtained by density functional
theory: $\left(\frac{\partial E_{p}}{\partial V}\right)_{S}=\frac{1}{N}\sum_{n<N}P^{DFT}\left(\left\{ u_{i\alpha}^{n}\right\} \right)$,
thus taking into account the anharmonicity of the potential. For the
derivative of the kinetic energy $E_{k}$, we use the quasi-harmonic
expression: $\left(\frac{\partial E_{k}}{\partial V}\right)_{S}=-\frac{1}{2V}\sum_{m}\hbar\omega_{m}\gamma_{m}\left(\frac{1}{2}+n_{B}\left(\omega_{m}\right)\right)$
\footnote{This expression comes from the equipartition of energy in harmonic
oscillators: $\left(\frac{\partial E_{k}}{\partial V}\right)_{S}=\frac{1}{2}\left(\frac{\partial E}{\partial V}\right)_{S}=\frac{1}{2}\left(\frac{\partial F}{\partial V}\right)_{T}$
and the usual quasi-harmonic derivation.}, with $\gamma_{m}$ the Gr\"uneisen parameter of mode $\omega_{m}$.
We then update the lattice parameter until the mean external pressure
becomes negligible. This allows us to calculate the thermal expansion
of the material, including even order anharmonic effects within the
effective 2nd order force constants, as shown for instance in Refs.\,\cite{Hellman_TDEP_potentials,Lan_TiO2_quarticity}.

\begin{figure}
\begin{centering}
\includegraphics[width=8.5cm]{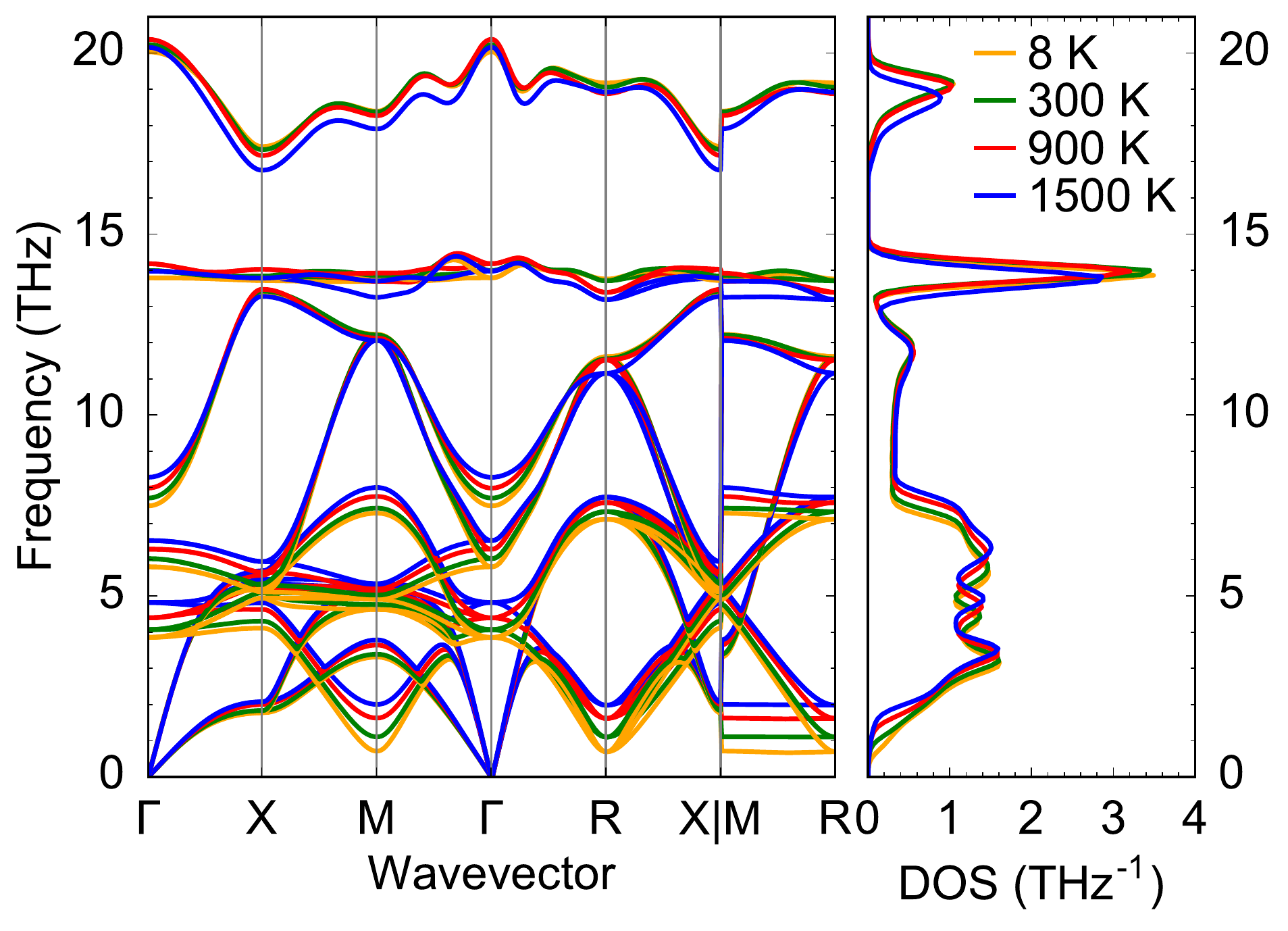} 
\par\end{centering}

\begin{centering}
\includegraphics[bb=0bp 0bp 512bp 223bp,width=8.5cm]{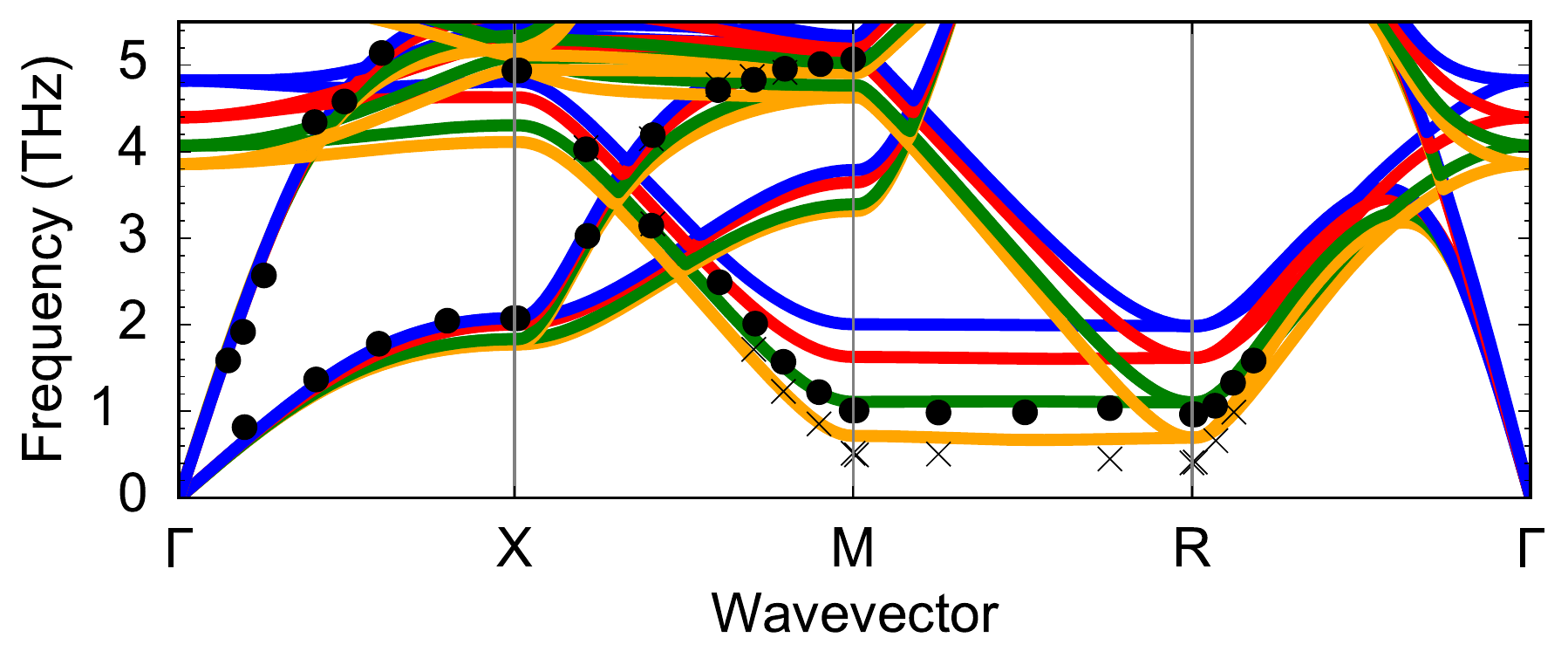} 
\par\end{centering}

\caption{(color online). Temperature-dependent phonon dispersion (left and
bottom) compared to the IXS measurements of Ref.\,\cite{Handunkanda_ScF3_QPT}
at 8\,K (black crosses) and 300\,K (black dots), and corresponding
density of states (right).\label{fig:Phonon_dispersion}}
\end{figure}

Fig.\,\ref{fig:Phonon_dispersion} shows the phonon spectra and density
of states obtained at 8\,K, 300\,K, 900\,K and 1500\,K compared
to the IXS measurements of the phonon frequencies at 8\,K and 300\,K
from Ref.\,\cite{Handunkanda_ScF3_QPT}. We observe that the effects
of temperature are antagonistic depending on the frequency range:
lower frequency modes harden with increasing temperature, while higher
frequency modes soften, with a boundary between those two regimes
at around 10\,THz. This is in qualitative agreement with the neutron-weighted
data of Ref.\,\cite{Li_ScF3_quartic}, except for the peaks around
3\,THz and 12\,THz, which were seen to have the opposite behavior.
Since both the phonon dispersions and the IXS data indeed point towards
a hardening of the lower frequencies, the disagreement with the neutron
weighted density of states likely comes from the phonon-neutron matrix
elements.

\begin{figure}
\begin{centering}
\includegraphics[width=8.5cm]{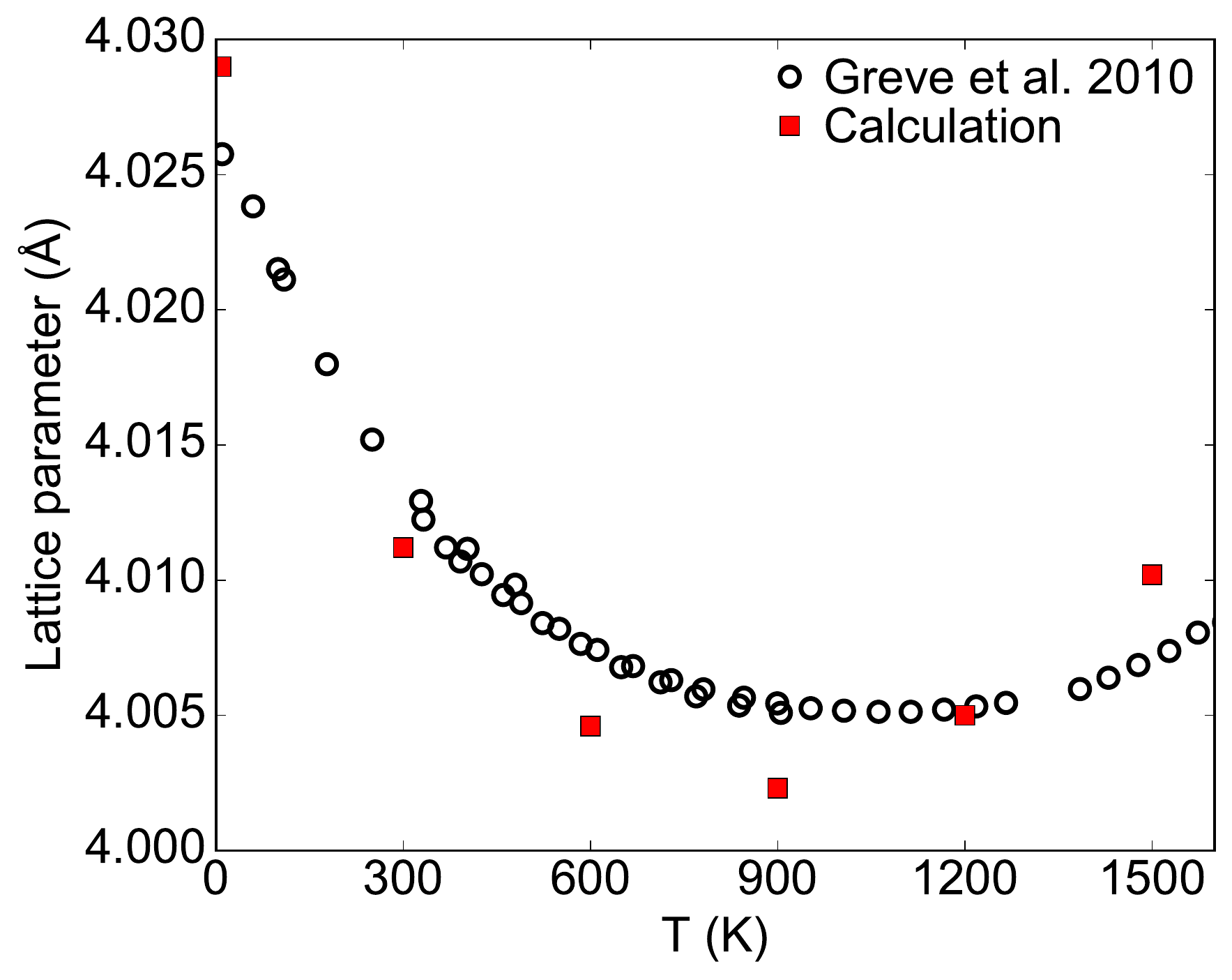} 
\par\end{centering}

\caption{(color online). Temperature-dependence of the lattice parameter of
ScF$_{3}$ using the PBEsol exchange-correlation potential compared
to the experimental data of Ref.\,\cite{Greve_ScF3_2010}.\label{fig:Lattice_parameter}}
\end{figure}

Fig.\,\ref{fig:Lattice_parameter} shows the calculated lattice parameter
compared to the experimental measurements of Greve and collaborators
\cite{Greve_ScF3_2010}. We use the PBEsol functional, which has been
conceived to give more reliable results for the crystal structure
of solids as compared to LDA or GGA \cite{Perdew_PBEsol}, and we
average the value of the lattice parameter over 30 cycles. We also
take into account electronic excitations in the thermal expansion
by using Fermi-Dirac statistics to determine the partial occupancies.
Indeed, electronic excitations have been shown to have a non-negligible
impact on the free energy, phonon dispersion and CTE in aluminium
and rhodium, in particular near the melting temperature \cite{Grabowski_Al,Grabowski_Thesis_2009}.
We obtain quantitative agreement up to 1500\,K, with a deviation
of the lattice parameter of less than 1\textperthousand.

\begin{figure}
\begin{centering}
\includegraphics[width=8.5cm]{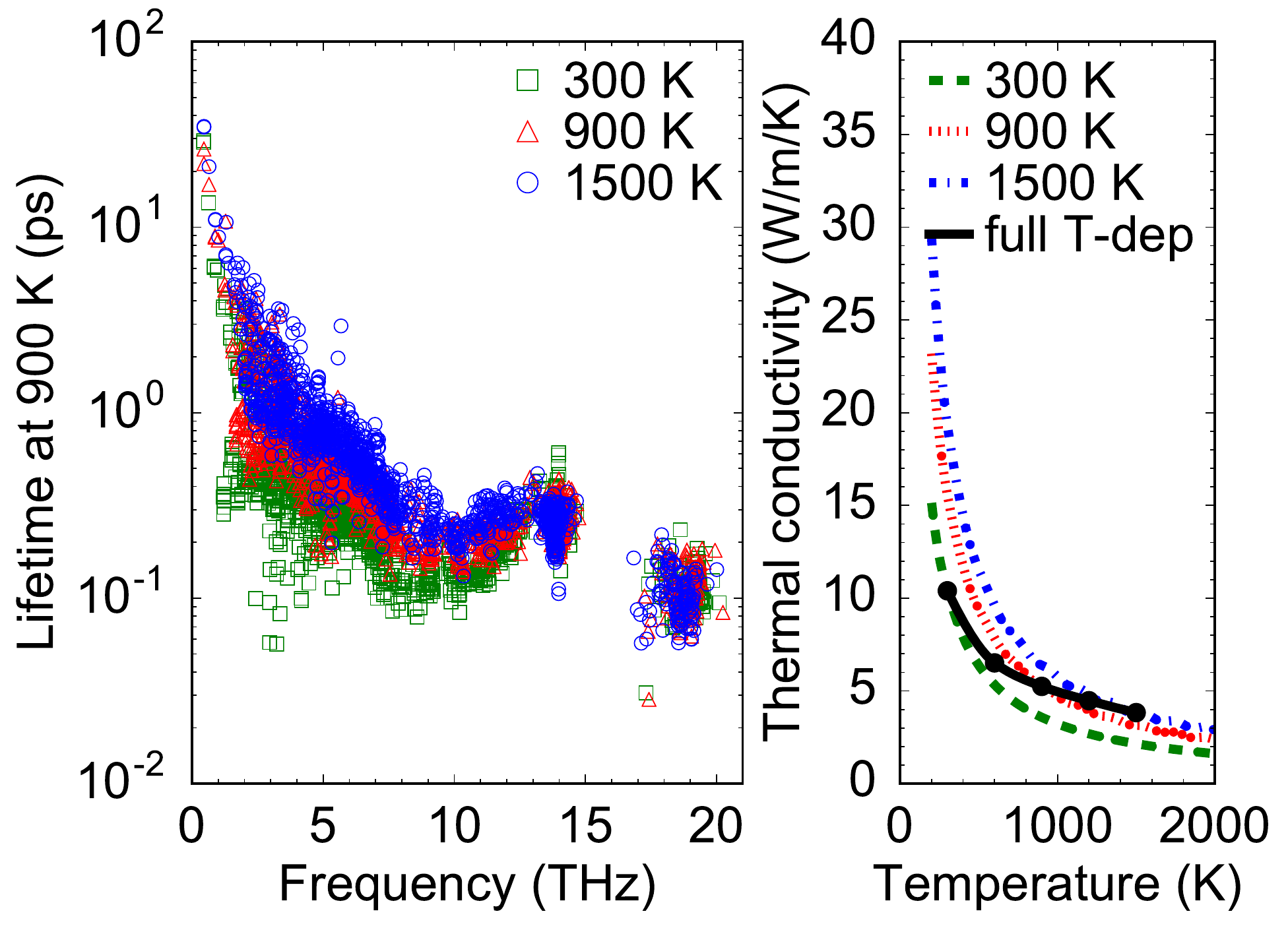} 
\par\end{centering}

\caption{(color online). Lifetimes of a 15x15x15 grid as a function of energy
at 900 K using force constants from different temperatures (left)
and thermal conductivity interpolated between calculations with fixed
2nd and 3rd order force constants corresponding to different temperatures
(right).\label{fig:Thermal_conductivity}}
\end{figure}

We now compute the thermal conductivity of ScF$_{3}$, which has neither
been measured nor previously calculated, up to our knowledge. We use
the full solution of the Boltzmann transport equation as implemented
in the ShengBTE code \cite{ShengBTE_2014}, and compute values from
300\,K to 1500\,K using the temperature-dependent effective 2nd
and 3rd order force constants. Fig.\,\ref{fig:Thermal_conductivity}
displays the lifetimes calculated at 900\,K while artificially using
force constants obtained for different temperatures, and the thermal
conductivity. When temperature increases, two important phenomena
occur simultaneously: the lifetimes of the soft modes become longer
due to the modification of their frequencies, but also all lifetimes
are enhanced globally. Indeed, close to a displacive instability the
three-phonon scattering rate of the soft mode becomes particularly
high because it is proportional to $\omega^{-1}$, as seen for instance
in SrTiO$_{3}$ \cite{Tani_soft_mode_lifetime}. The global enhancement
of lifetimes is mainly due to the reduction of the available phase
space for three-phonon processes \cite{Lindsay_phase_space}\footnote{In passing, we note that the modification of the lattice parameter
in itself has nearly no incidence on the thermal conductivity.}, which in turn is due to the shrinking in energy of the phonon density
of states by the simultaneous softening of high energy modes and hardening
of low energy modes. As a result, the thermal conductivity acquires
an anomalous temperature-dependence, that we can roughly describe
by a power-law close to $\kappa\varpropto T^{-0.6}$ (see Fig.\,\ref{fig:Thermal_conductivity}).
This behavior is reminiscent of what has been found in other perovskites,
such as the prototype compound SrTiO$_{3}$ \cite{Muta_SrTiO3_kappa,Popuri_SrTiO3_kappa},
in which the thermal conductivity has been shown to follow a power
law $\kappa\varpropto T^{-\alpha}$ with $\alpha\thickapprox0.6-0.7$.
We speculate that the presence of soft modes in many perovskites --
which are very sensitive to temperature -- makes such anomalous dependence
a common behavior in this class.

To better understand the suppression of NTE in ScF$_{3}$, let us
begin by summarizing the problem in simple arguments. Within the quasi-harmonic
approximation, the sign of the CTE $\alpha_{V}$ follows the weighted
Grüneisen parameter $\gamma$ via $\alpha_{V}=\frac{\gamma C_{v}\rho}{K_{T}}$,
with $K_{T}$ the isothermal bulk modulus, $C_{v}$ the isochoric
heat capacity and $\rho$ the density \cite{Gruneisen,Ashcroft_Mermin}.
In this material, all phonon modes have energies lower than 1000\,K,
which means that the heat capacity and weighted Grüneisen parameter
have negligible thermal variation above this temperature. Moreover,
the volume of the unit cell at 1300\,K and 900\,K is approximately
the same, excluding the possibility that the potential volumic variation
of the Grüneisen parameter could explain by itself the change of expansion
regime. It is thus very unlikely that the suppression of NTE around
1100\,K can be explained within the quasi-harmonic approximation.
This is in contrast to the other famous compound exhibiting NTE over
a large temperature range, ZrW$_{2}$O$_{8}$, in which NTE persists
until its decomposition \cite{Mary_ZrW2O8_NTE}; and to many compounds
which exhibit NTE over a small temperature range before its suppression
due to the thermal population of higher-energy phonon modes, such
as the prototype material of the empty perovskite family ReO$_{3}$
\cite{Chatterji_ReO3_NTE,Dapiaggi_ReO3_NTE} or, more prosaically,
silicon \cite{Gibbons_Si_NTE}. Thus, ScF$_{3}$ presents an explicit
case where high-order anharmonicity plays a crucial role. In passing,
we note that even in those compounds with apparently simpler behavior
there remain important related questions, such as the possible role
of quartic anharmonicity in the temperature-dependence of the phonon
spectrum of ZrW$_{2}$O$_{8}$ \cite{Ernst_ZrW2O8_gruneisen}; or
the reappearance of NTE in ReO$_{3}$ around 650\,K \cite{Chatterji_ReO3_extended}
which could not be explained within the Grüneisen formalism \cite{Chatterji_ReO3_NTE,Wdowik_ReO3_lattice_dynamics}
and was tentatively attributed to the strong anharmonic behavior of
a soft mode \cite{Chatterji_ReO3_extended,Chatterji_ReO3_phonon_Tdep}.

We now discuss the CTE in terms of the Grüneisen parameters \footnote{Such a discussion stays meaningful in the present case, since the
definition $\alpha_{V}=\frac{\gamma C_{v}\rho}{K_{T}}$ gives $\gamma=\frac{1}{\rho C_{v}}\left(\frac{\partial S}{\partial V}\right)_{T}$,
the thermal density matrix $\rho_{h}$ is calculated within the harmonic
approximation and $S=-k_{B}\text{Tr}\left(\rho_{h}\ln\rho_{h}\right)$.}. We calculate the mode-dependent Grüneisen parameters $\gamma_{m}=-\frac{V}{\omega_{m}}\left(\frac{\partial\omega_{m}}{\partial V}\right)_{T}$
and the weighted Grüneisen parameter $\gamma=\sum_{m}\gamma_{m}c_{vm}/\sum_{m}c_{vm}$,
using the 2nd and 3rd order force constants of a given temperature
as \cite{Fabian_Si_expansion_Gruneisen,Broido_Si_kappa,Hellman_TDEP_2013}:
\[
\gamma_{m}=-\frac{1}{6\omega_{m}^{2}}\sum_{ijk\alpha\beta\gamma}\frac{\epsilon_{mi\alpha}^{*}\epsilon_{mj\beta}}{\sqrt{M_{i}M_{j}}}r_{k}^{\gamma}\Psi_{ijk}^{\alpha\beta\gamma}e^{i\mathbf{q}\cdot\mathbf{r}_{j}}
\]
Since the sign of the CTE $\alpha_{V}=\frac{\gamma C_{v}\rho}{K_{T}}$
is the same as $\gamma$, the evolution of $\gamma$ with temperature
allows us to interpret the change of regime in terms of the evolution
of the contribution of different phonon modes. In the high-temperature
limit and if we do not take into account the modification of the 2nd
and 3rd order force constants, the weighted Grüneisen parameter becomes
constant and equal to the arithmetical mean of all mode-dependent
Grüneisen parameters.

\begin{figure}
\begin{centering}
\includegraphics[width=8.5cm]{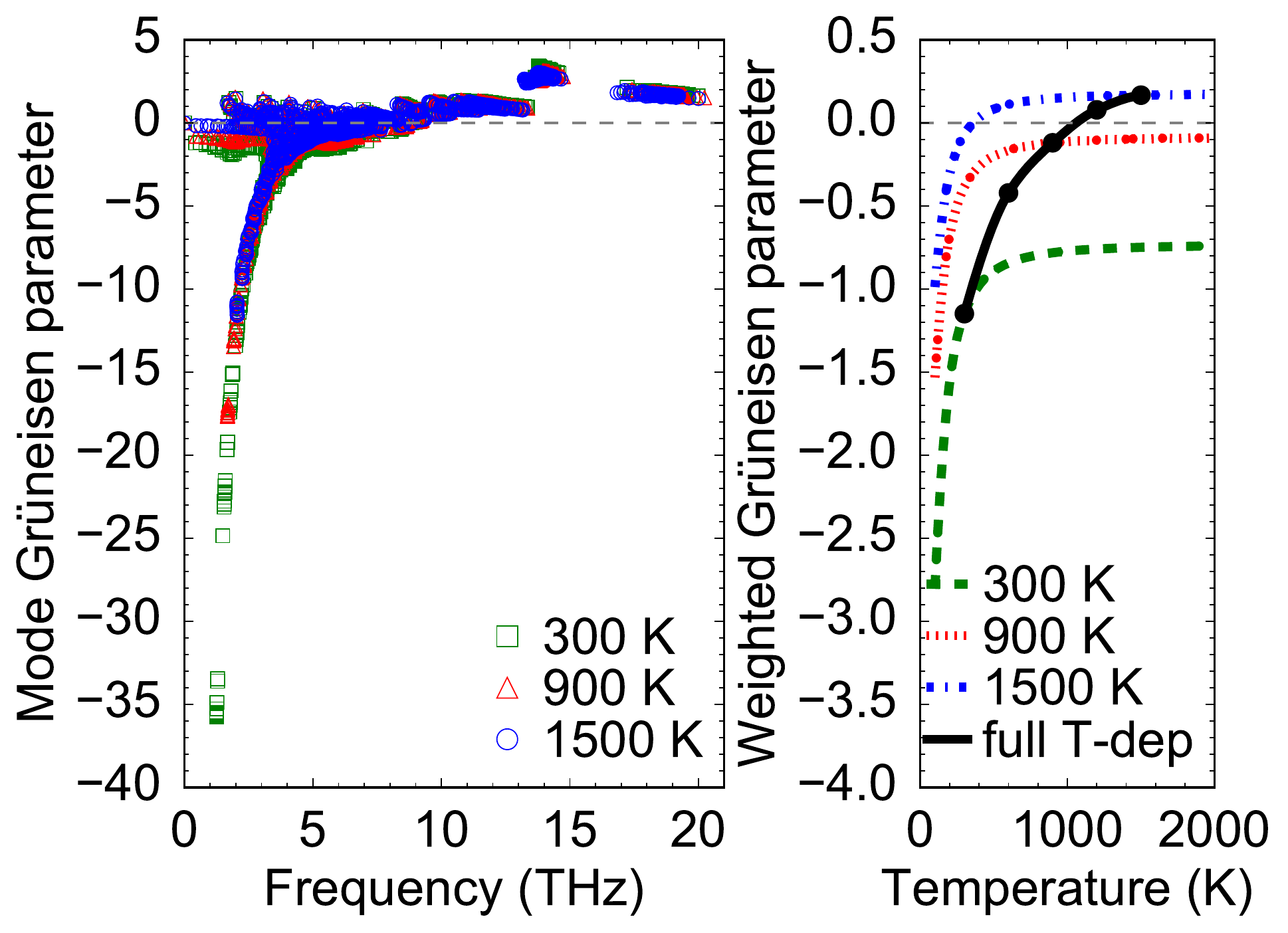} 
\par\end{centering}

\caption{(color online). Mode-dependent Grüneisen parameters of a 15x15x15
grid as a function of energy and temperature (left) and weighted Grüneisen
parameter interpolated between calculations with fixed 2nd and 3rd
order force constants corresponding to different temperatures (right).\label{fig:Gruneisen}}
\end{figure}

In Fig.\,\ref{fig:Gruneisen}, we decouple the effect of temperature
due to the modification of the weight of the mode heat capacity from
the effect due to the modification of the force constants by tracing
several temperature-dependent weighted Grüneisen parameter curves
using fixed force constants obtained at different temperatures. A
large contribution to the NTE comes from the soft mode line between
the R and M points, due to its small frequency and large variation
-- which shows that this type of compounds close to a mechanical instability
are good candidates for NTE. When the temperature is increased, these
mode Grüneisen parameters are mostly impacted by the modification
of their frequency, which lowers their contribution to NTE. Still,
the mode Grüneisen parameters at a given frequency are globally pushed
up by temperature in the low-energy region. This is similar to what
happens for the lifetimes, as discussed above. Both effects are necessary
to quantitatively understand the suppression of NTE that appears around
1100\,K (see right panel of Fig.\,\ref{fig:Gruneisen}), in agreement
with the experiment.

It is also remarkable that the full temperature dependence of the
weighted Grüneisen parameter differs importantly from the one obtained
within the quasi-harmonic approximation in the whole temperature range,
showing that higher-order anharmonic effects are crucial in the behavior
of the CTE even at lower temperatures. This last statement is all
the more true if one takes into account the variation of the volume
to recalculate the phonon dispersions. Indeed, the soft mode frequencies
are lowered when the lattice parameter is reduced and a phase transition
can take place under pressure \cite{Aleksandrov_ScF3_pressure,Handunkanda_ScF3_QPT}.
In the standard Grüneisen formalism, this translates in a lowering
of the phonon frequencies with temperature via $(\partial\omega_{m}/\partial T)_{P}=(\partial\omega_{m}/\partial V)_{T}(\partial V/\partial T)_{P}=-\omega_{m}\gamma_{m}\alpha_{V}$.
In contrast, our calculations and the experimental data show that
the soft mode frequencies \textit{increase} with temperature. A similar
anomaly has been observed in experimental measurements of the phonon
density of states in ZrW$_{2}$O$_{8}$ \cite{Ernst_ZrW2O8_gruneisen}
and tentatively attributed to quartic anharmonicity, which is indeed
partially captured in our method. Interestingly, the same anomalous
behavior happens also for the higher energy optical modes, that are
softened while the volume decreases in spite of a positive Grüneisen
parameter, showing that anharmonicity plays an important role over
the whole phonon spectrum.

In conclusion, we have used temperature-dependent anharmonic calculations
to compute the evolution of the phonon frequencies and scattering
rates in ScF$_{3}$. We have predicted an anomalous temperature dependence
of the thermal conductivity, notably due to the hardening of the soft
modes frequencies with increasing temperature, and interpreted it
as a general feature of this class of compounds. Finally, we have
shown that the weighted Grüneisen parameter acquires a dependence
in temperature that is qualitatively different from the quasi-harmonic
behavior and explains the suppression of negative thermal expansion.
This demonstrates that taking into account high-order anharmonicity
is crucial for the qualitative and quantitative computation of physical
properties at intermediate and high temperatures. Furthermore, this
work paves the way towards the search for new materials displaying
negative thermal expansion or low thermal conductivity from \textit{ab-initio}
calculations over a large temperature range.
\begin{acknowledgments}
We acknowledge useful discussions with Ion Errea, Ole Hellman, Wu
Li and Stefano Curtarolo. This work is partially supported by the
French ``Carnot'' project SIEVE.
\end{acknowledgments}

\end{document}